\documentclass[11pt]{article}
\usepackage{hyperref}
\pdfoutput=1

\begin{document}
\title{Vortex shedding and hovering of a rigid body in an oscillating flow}
\author{Bin Liu$^1$, Annie Weathers$^2$, Stephen Childress$^1$, and Jun Zhang$^{1,2}$\\
\\\vspace{6pt} $^1$Applied Mathematics Laboratory, Courant Institute of Mathematical Sciences, 
\\\vspace{6pt}New York University, New York, NY 10012, USA
\\ $^2$Department of Physics, New York University, New York, NY 10003, USA}
\maketitle
\begin{abstract}
The fluid dynamics video shows rigid, spatially asymmetric bodies interacting with oscillating background flows. A free rigid object, here a hollow ``pyramid," can hover quite stably against  gravity in the oscillating airflow with a zero mean, when its peak speed is sufficiently high. We further show in shadowgraph imaging how this asymmetric body sheds vortices in such an  unsteady flow, thus enabling the body to ``ratchet" itself  through the background flow.

\end{abstract}
\section{Introduction}

In the videos (\href{http://ecommons.library.cornell.edu/bitstream/1813/14101/2/mpeg-1.mpg}{low-res.} and
\href{http://ecommons.library.cornell.edu/bitstream/1813/14101/3/mpeg-2.mpg}{high-res.}), we first present an example of free hovering of a rigid, spatially asymmetric body in an oscillating airflow. The rigid object is a hollow ``pyramid"  of height $h=3.2$ cm high, constructed of carbon fibers and wax paper. The background air oscillates at a frequency $f=20$ Hz. When the air speed is high enough ($V_\textrm{max}=1.7$ m/s), the paper pyramid hovers against its weight ($W=284$ dynes, or $m=0.29$ g).  The pyramid aligns itself spontaneously in the lifting position, exhibiting surprising stability. The video is recorded by a high speed camera and then played back at 1/13 of the real speed. 

In the second part  of the video, we show the vortical structures produced by the interaction between a  ``V"-shaped object and an oscillating background flow. The flow visualization is realized by shadowgraph imaging of an oscillating water flow around a two-dimensional pyramid of a height $h=3.0$ cm, composed of two faces. In the example, shown in real-time in the video, the oscillating frequency of the water flow is $0.8$ Hz, with an amplitude of 2 cm. The system thus has approximately the same Reynolds number regime as the experiments in air. Vortices generated during each period coalesce into paired vortices, before  detaching and propagating downward, producing the momentum flux needed to support the body.
\end{document}